\documentclass[twocolumn,showpacs,preprintnumbers,prl,amsmath,amssymb]{revtex4}%PRL
\usepackage{graphicx}% Include figure files
\usepackage{dcolumn}% Align table columns on decimal point
\usepackage{bm}% bold math
\usepackage{longtable}% tables which need more space than one page
\usepackage{array}
\usepackage{color}
\usepackage{hyperref}
\begin{document}

% \preprint{APS/123-QED}
\title{Photoluminescence Spectroscopy of the Molecular Biexciton in Vertically Stacked Quantum Dot Pairs}
\author{M. Scheibner}
\email{scheibner@bloch.nrl.navy.mil}
\author{I. V. Ponomarev}
\author{E. A. Stinaff}
\author{M. F. Doty}
\author{A. S. Bracker}
\author{C. S. Hellberg}
\author{T. L. Reinecke}
\author{D. Gammon}
\affiliation{Naval Research Laboratory, Washington, DC 20375, USA
}

\date{\today}% It is always \today, today,
             %  but any date may be explicitly specified
\begin{abstract}
We present photoluminescence studies of the molecular neutral
biexciton-exciton spectra of individual vertically stacked
InAs/GaAs quantum dot pairs. We tune either the hole or the
electron levels of the two dots into tunneling resonances. The
spectra are described well within a few-level, few-particle
molecular model. Their properties can be modified broadly by an
electric field and by structural design, which makes them highly
attractive for controlling non-linear optical properties.
\end{abstract}
\pacs{78.67.Hc, 73.21.La, 78.55.Cr}% PACS, the Physics and Astronomy
                             % Classification Scheme.
                             %\keywords{Suggested keywords}%Use showkeys class option if keyword
                              %display desired
\maketitle
\begin{figure}[t]
\centering
\includegraphics[width=7cm]{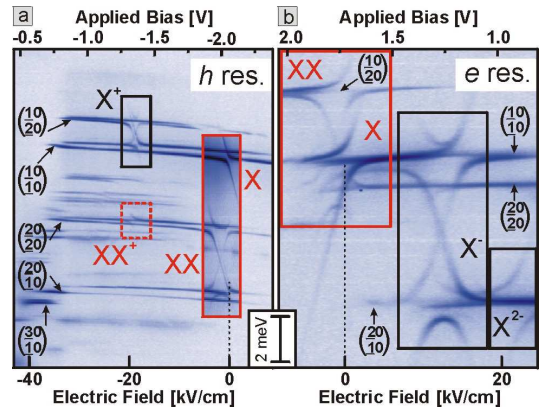}% Here is how to import EPS art
\caption{\label{fig:Overview} (Color online) \textbf{(a)} Hole and
\textbf{(b)} electron level resonances in the field dependent PL
spectra of a QDM. The red boxes highlight the regions were the
molecular resonances of exciton, biexciton and charged biexciton
are seen. The area in the solid red box in (a) was taken with 5
times higher resolution in the field direction and 8 times longer
integration time. The black boxes outline the patterns of $X^+$,
$X^-$ and $X^{2-}$, which were previously studied
\cite{StinaffScience06, Krenner06, ScheibnerPRB07}. The spectra
are centered at (a) 1251.57 meV and (b) 1297.21 meV.}
\end{figure}

% ***************************************************************
% *  Introduction                                               *
% ***************************************************************
\indent Quantum dots (QDs) provide an attractive active medium for
nanophotonics, quantum optics and quantum information because they
can be integrated into opto-electronic architectures, they
interact strongly with light, and their properties are adjustable
by growth.  This functionality can be enhanced by building
molecules of QDs (QDMs), in which new inter-dot transitions arise
that are strongly tuned with electric field. Here we employ
single, custom designed "diatomic" QDMs in order to explore the
electric field response of a new type of optical excitation in
QDMs - the molecular biexciton.
\\
\indent The exciton and biexciton states are at the heart of many
quantum and nonlinear optical processes in a single QD
\cite{SantoriPRB02}\nocite{StevensonNature06, AkopianPRL06,
LiScience03}-\cite{ReimerQuantPhys07}. Their energy level diagram
is the starting point in all such studies. The energy of the
biexciton is not exactly twice that of the exciton because of
Coulomb and spin interactions. In a QDM these interactions again
play a central role in the energy spectrum, but now intricately
combined with coherent tunneling between dots. The resulting
molecular biexciton spectra are very rich and sensitive to
electric field, but their characteristic spectral pattern can be
understood quantitatively using a relatively simple energy level
diagram. The spectral patterns presented here are qualitatively
similar for all QDMs that we have studied as long as we account
for whether it is the electrons or the holes that tunnel between
dots.  We demonstrate both cases here. In this paper we consider
neutral QDMs but a similar analysis can be applied to charged
QDMs.
\\
\indent The QDM easiest to fabricate, and the one now under
intense study, is the asymmetric diatomic molecule formed out of
two non-identical InAs dots vertically stacked and separated by a
thin GaAs tunnel barrier. In this study we used QDMs with tunnel
barriers of $d=6$ nm and $d=14$ nm (center to center
$\tilde{d}=8.5$ nm and 17 nm) for hole and electron tunneling
respectively. They were embedded in Schottky diodes. We chose the
heights of the QDs such that the top QD had higher transition
energy. Here we consider only the lower energy QD transition - the
other is at substantially higher energy and plays no role. Further
details on the growth and experimental methods can be found in
Refs. \cite{ScheibnerPRB07, BrackerAPL06}.
\\
% ***************************************************************
% *  Figure 1                                                   *
% ***************************************************************
\begin{figure}[b]
\includegraphics[width=7cm]{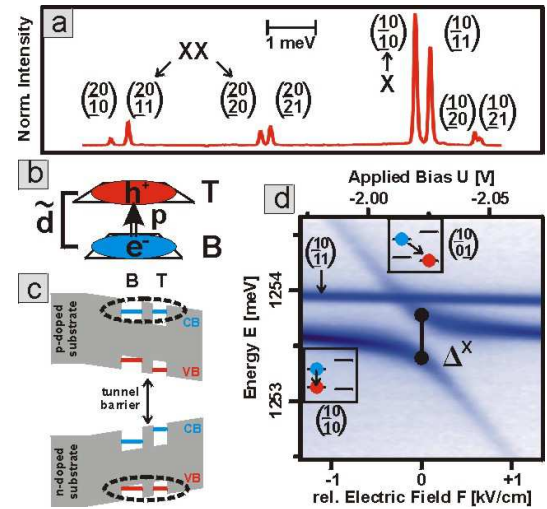}% Here is how to import EPS art
\caption{\label{fig:neutralexciton}(Color online) \textbf{(a)} PL
spectrum of a QDM showing bottom dot intra-dot transitions
($U=-1.60$ V). \textbf{(b)} A large dipole moment $p\propto
e\tilde{d}$ exists between the electron and hole, each located in
separate QDs. \textbf{(c)} Electric field dependent PL spectrum at
the hole level resonance of the neutral exciton obtained from a
QDM with $\tilde{d}=8.5$ nm. An AC is seen at the resonance of
intra-dot ($^{\underline{1}0}_{\underline{1}0}$) and inter-dot
($^{\underline{1}0}_{0\underline{1}}$) transition, arising from
the formation of bonding and anti-bonding molecular hole states
with a splitting of $\Delta^X=420$ $\mu$eV. \textbf{(d)}QDM diode
structures for electron (top) and hole (bottom) level resonances
(B: bottom QD, T: top QD). }
\end{figure}
\indent A typical optical spectrum of the bottom dot transitions
in a single QDM exhibits a series of PL-line doublets (Fig.
\ref{fig:neutralexciton}(a)). The lower energy lines in the
doublets form a sequence of charged excitonic transitions, which
is similar to that measured in single dots; i.e. $X^-$
($^{\underline{2}0}_{\underline{1}0}$), $XX$
($^{\underline{2}0}_{\underline{2}0}$), $X$
($^{\underline{1}0}_{\underline{1}0}$), and $X^+$
($^{\underline{1}0}_{\underline{2}0}$). Here the upper two numbers
are the numbers of the electrons in the bottom and the top dot,
respectively, and the lower two numbers are the numbers of holes
in the two dots. The underlines denote the position of the
recombining particles (in this case, identifying an intra-dot
transition). The shifts in the energies between the differently
charged exciton transitions are well known in the spectra of
single dots and arise from quantitative differences in the Coulomb
interactions of the holes and electrons. The second line in each
doublet is a new transition characteristic of a QDM. It is a Stark
shifted replica of the first line, caused by the electric field of
an extra charge in the top dot (here one hole)
\cite{PL-triplets2}. Note in particular the new $X^-$-like
biexciton transition ($^{\underline{2}0}_{\underline{1}1}$). These
assignments are solidified below. This energy structure serves as
a useful basis for the interpretation of the full and often
complex molecular spectrum.
\\
\indent A QDM allows for two types of optical transitions -- (i)
the intra-dot transition mentioned above in which only one QD is
involved, and (ii) the inter-dot transition in which different QDs
are involved. For the intra-dot exciton transition, as for single
dots, the static dipole moment (experimentally measured by the
shift of the transition energy with electric field) is fairly
small. In contrast, for the inter-dot transition the static dipole
moment is very large because of the separation, $\tilde{d}$,
between the dots ($p\propto e\tilde{d}$ where $e$ is the electron
charge) (Fig. \ref{fig:neutralexciton}(b)).
\\
\indent The molecular nature of these QD pairs is revealed in the
coherent superposition of states that arises when the electric
field brings the intra-dot and inter-dot exciton states
energetically into resonance. Because the two dots in the QDM have
different transition energies, either the electron or the hole
(within the exciton) can tunnel, but not both (Fig.
\ref{fig:neutralexciton}(c)) \cite{BrackerAPL06}.  This has been
shown for a single exciton in neutral and in charged QDMs
\cite{Krenner05}\nocite{OrtnerPRL05,StinaffScience06}-\cite{Krenner06}
(see also Fig. \ref{fig:Overview}). We consider first the case in
which the hole levels of the two dots are near resonance, and the
two electron levels are detuned far from resonance. Coherent
tunneling of holes at these resonances leads to an anticrossing
(AC) in the energy levels and to a mixing of the properties of the
exciton states. For the exciton ($X$), a single AC of energy
$\Delta^X=430$ $\mu$eV is seen in the PL spectrum (Fig.
\ref{fig:neutralexciton}(d)).
\\
% ***************************************************************
% *  Fig. States XX-Cascade                                       *
% ***************************************************************
% ***************************************************************
% *  Fig. 1                                                     *
% ***************************************************************
\begin{figure}[t]
 \centering
\includegraphics[width=7cm]{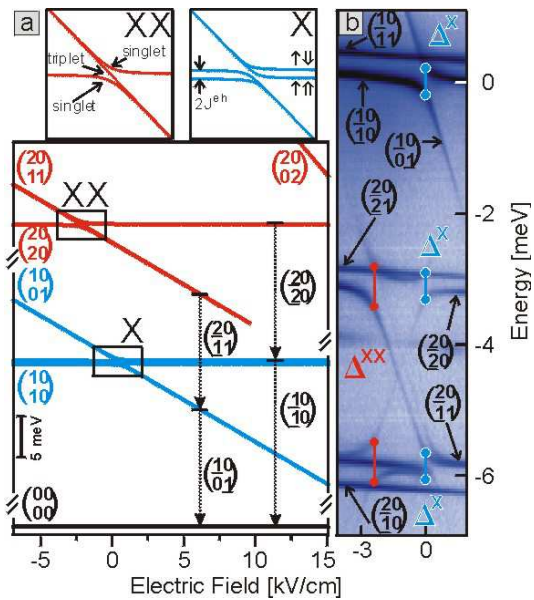}% Here is how to import EPS art
\caption{\label{fig:XXcascadeHtype}(Color online) \textbf{(a)} The
calculated energy level diagram of the neutral biexciton-exciton
cascade \cite{Z-HamiltonX0-alt, Z-ValuesXXHamilton-htype}. The
areas in the boxes $X$ and $XX$ have been enlarged to show the
fine structure caused by spin exchange interactions in the
vicinity of level resonances. \textbf{(b)} Extended electric field
dependent optical spectrum of the QDM (same as in Fig.
\ref{fig:neutralexciton}(d)) with neutral exciton ($X$) and
neutral biexciton ($XX$) transitions.}
\end{figure}
\indent If two electron-hole pairs are excited in the system, a
biexciton ($XX$) is formed. The biexciton state allows a
(conditional) two photon cascade - optical transitions from the
biexciton state to the exciton states, and from the exciton states
to the ground state (exciton vacuum). In a single dot, the lowest
energy state of the biexciton is a spin singlet with two electrons
and two holes, each spin-paired in s-shell orbitals.  As one might
expect, the quantum states of a molecular biexciton are
significantly richer than in a single QD because the carriers can
be distributed over both dots. Nevertheless, as we now show, the
regular pattern found experimentally, and from analysis, leads to
a simple, intuitive understanding of the molecular biexciton.
\\
\indent Figure \ref{fig:XXcascadeHtype}(a) shows the calculated
energies for the biexciton ($XX$) and the exciton ($X$)
\cite{Z-HamiltonX0-alt, Z-ValuesXXHamilton-htype}. The biexciton
can take several configurations of the two electrons and two holes
over the two dots.  In the hole-resonance case, both electrons
relax to the s-shell orbital of the low energy dot, and thus form
an electron spin singlet configuration by the Pauli Principle. The
two holes can be in either of the two dots.  We calculate energies
relative to that configuration in which all particles are in the
low energy dot. Thus, with all electrons and holes in the same
dot, ($^{20}_{20}$), the slope as a function of electric field is
zero.  With only one hole in the other dot, ($^{20}_{11}$), the
slope is $p$.  With two holes in the other dot ($^{20}_{02}$), the
slope is $2p$. For the configuration in which the two holes are
separated from the two electrons ($^{20}_{02}$), there is a large
Coulomb interaction that substantially increases its energy.
\\
\indent Molecular features in the biexciton spectrum appear along
with the exciton spectrum in Fig. \ref{fig:XXcascadeHtype}(b),
where we present a new view of the data of Fig.
\ref{fig:neutralexciton}(d) over a wider energy range. There are
several observations that allow us to identify, and understand
this spectrum and to define fitting parameters that can largely be
found independently from each other and need to be fine tuned only
for quantitative agreement.
\\
\indent \textbf{(i)} There is an ``X-pattern'' that ranges between
two extremes in energy, that are determined by the biexciton
\textit{intra}-dot transitions in Fig.
\ref{fig:XXcascadeHtype}(b). On the high energy side, the
``X-pattern'' is bounded by a nearly horizontal line that
corresponds closely to the previously known biexciton transition
energy of a single dot. Thus, we were able to conclude that this
is the intra-dot transition
($^{\underline{2}0}_{\underline{2}0}$). On the low energy side of
the ``X-pattern'', the transition is $X^-$-like, which means,
close to the transition energy of a negative exciton in the bottom
dot ($^{\underline{2}0}_{\underline{1}0}$). Thus, we deduce that
this low energy line is the intra-dot transition
($^{\underline{2}0}_{\underline{1}1}$). As mentioned above, it is
shifted from the negative exciton transition because of the
presence of a second hole in the other dot.
\\
\indent \textbf{(ii)} Within the ``X-pattern'' there are two
inter-dot transitions that anticross with the two intra-dot
transitions. The ``X-pattern'' observed in the data occurs because
the biexciton and exciton states each have energy ACs that are
close in electric field (because of the relatively small
differences in Coulomb energies for the states involved). One pair
of these AC resonances occurs at precisely the same electric field
as the single exciton resonance. This electric field alignment
happens because the biexciton makes a transition to the exciton
state, and thus the AC of the exciton state appears in the optical
transitions of both the exciton and biexciton. Consequently, the
ACs on the left side of the biexciton spectrum in Fig.
\ref{fig:XXcascadeHtype}(b) must arise from the biexciton level
ACs.  The AC energy for the biexciton levels is $\Delta^{XX}=630$
$\mu$eV $\approx\sqrt{2}\Delta^{X}$ because two indistinguishable
holes can tunnel \cite{StinaffScience06}.
\\
\indent \textbf{(iii)}  The distinctive fine structure splittings
in the biexciton spectrum, which are best seen at the two
energetically lowest ACs in Fig. \ref{fig:XXcascadeHtype}(b),
arise from spin.  The biexciton states in which both holes are in
the same dot, ($^{20}_{20}$) and ($^{20}_{02}$), are spin
singlets.  If the two holes are each in a separate dot,
($^{20}_{11}$), the biexciton can exist in a spin singlet or in
one of three spin triplet states.  Because tunneling conserves
spin, singlet states anticross only with singlet states. In
essence, Pauli blocking prevents the triplets from tunnel-coupling
with a singlet. Thus, the triplet states pass straight through the
biexciton AC, and there is a ``kinetic exchange splitting'' of the
singlet and triplet state energies that arises from tunneling and
Pauli blocking (top left inset of Fig.
\ref{fig:XXcascadeHtype}(a)). Moreover, the additional fine
structure splitting in the exciton \cite{ScheibnerPRB07}, arising
from exchange of the unpaired electron and hole ($J^{eh}$) (top
right inset of Fig. \ref{fig:XXcascadeHtype}(a)), is seen clearly
in the transition spectrum between biexciton and exciton. This
fine structure does not appear in optical transitions from the
exciton state because one transition is optically dark.  These
exchange splittings in the calculated energy diagrams (Fig.
\ref{fig:XXcascadeHtype}(a)) can be traced to the corresponding
fine structure patterns observed in the measured spectrum (Fig.
\ref{fig:XXcascadeHtype}(b)). Thus, we find that our understanding
of the fairly complex spectral pattern of the biexciton is
described remarkably very well by the simple energy level diagrams
of Fig. \ref{fig:XXcascadeHtype}(a).
\\
% ***************************************************************
% *  Fig. Overview                                            *
% ***************************************************************
% ***************************************************************
% *  Fig. Theory compare                                            *
% ***************************************************************
\begin{figure}[b]
\centering
\includegraphics[width=8cm]{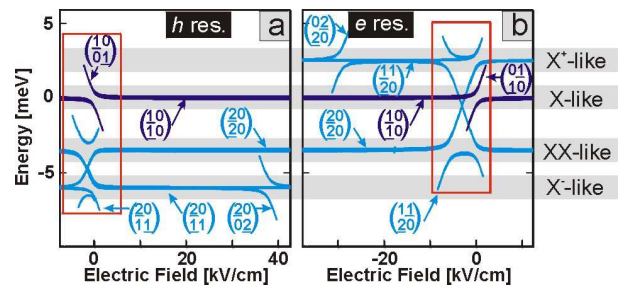}% Here is how to import EPS art
\caption{\label{fig:TheorySteve}(Color online) Calculated optical
spectra of the neutral excitons (blue) and the corresponding
biexcitons (cyan) if \textbf{(a)} the hole levels and \textbf{(b)}
the electron levels are tuned into resonance. For illustration we
have arbitrarily taken the same parameters for both cases -- in
meV: $E^{eh}=17.5$, $E^{h}_{1}=14.8$, $E^{e}_{1}=23.3$,
$J^{eh}=0.2$, $\Delta^X=1.0$, $p=\frac{\pm1}{(\text{kV/cm})}$
\cite{Z-HamiltonX0-alt, Z-ValuesXXHamilton-htype,
Z-ValuesXXHamilton-etype}.}
\end{figure}
\indent We can also engineer the asymmetric QDM structure to
induce the electron to tunnel instead of the holes
\cite{BrackerAPL06}. In Fig. \ref{fig:Overview} we compare the two
cases of hole and electron tunneling over a large bias range. We
focus on the biexciton-exciton cascade spectra in the solid red
boxes. The same discussion given above for the hole resonance
applies - with two major differences: (i) Because electrons
instead of holes now occupy the top QD in the inter-dot
configurations, e.g. ($^{01}_{10}$) instead of ($^{10}_{01}$), the
sign of the inter-dot dipole moment is reversed. (ii) The
biexciton ``X-pattern'' spectrum now overlaps the exciton
spectrum; in contrast to the hole resonance case, where it was
shifted well below in energy. This difference arises because of
the different nature of one of the direct transitions that bounds
the ``X-pattern''.  In particular, the
($^{\underline{1}1}_{\underline{2}0}$) direct transition that
occurs in the electron resonance case is now shifted above the
neutral exciton energy and close to the energy of a positive
exciton ($X^+$-like). This is in contrast to the corresponding
transition for the hole resonance case,
($^{\underline{2}0}_{\underline{1}1}$), whose energy was close to
the negative exciton ($X^-$-like) and at an energy much lower than
the exciton. For illustration, we compare calculated transition
spectra for the two cases in Fig. \ref{fig:TheorySteve}. The boxed
regions reproduce qualitatively the measured spectra (red boxes in
Fig. \ref{fig:Overview}) very well.
\\
\begin{figure}[t!]
\centering
\includegraphics[width=8cm]{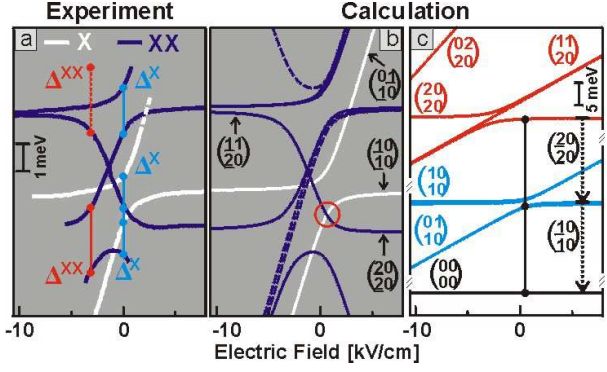}% Here is how to import EPS art
\caption{\label{fig:XXcascadeEtype}(Color online) \textbf{(a)}
Measured field dependent transition energies of the neutral
biexciton cascade from the red boxed region in Fig.
\ref{fig:Overview}(b) ($X$ (white) and $XX$ (blue) PL-lines).
\textbf{(b)} Fitted PL-spectrum \cite{Z-HamiltonX0-alt,
Z-ValuesXXHamilton-etype}. \textbf{(c)} The corresponding level
diagram.}
\end{figure}
\indent For a more quantitative analysis of the electron tunneling
case we focus on the area in the red box in Fig.
\ref{fig:Overview}(b) (see Figs. \ref{fig:XXcascadeEtype}(a) and
(b)). Because the electron mass is much smaller than the hole
mass, the AC energy is much larger (1.65 meV) than for the hole
resonance case (0.45 meV) shown above, even with a dot separation
twice as large. This leads to other quantitative differences in
the spectrum as well. In particular, relative intensities are
significantly changed because of more efficient thermalization and
differences in oscillator strengths, which prevent the observation
of some lines (dashed in Fig. \ref{fig:XXcascadeEtype}(b)).
\\
\indent The molecular cascade transitions that are similar to
those for a single dot, ($^{\underline{2}0}_{\underline{2}0}$) and
($^{\underline{1}0}_{\underline{1}0}$), are shown in Fig.
\ref{fig:XXcascadeEtype}(c) by the dashed vertical arrows at $F=6$
kV/cm \cite{Z-HamiltonX0-alt, Z-ValuesXXHamilton-etype}. With QDMs
the transition energies are strongly tunable with bias. For
example, moving to the left from $F=6$ kV/cm in Fig.
\ref{fig:XXcascadeEtype}(c), the exciton and biexciton transition
energies change continuously, reaching a field where the two
become equal \cite{ReimerQuantPhys07}. This point is shown by the
solid vertical lines in Fig \ref{fig:XXcascadeEtype}(c), and
corresponds to the circled crossing point in Fig.
\ref{fig:XXcascadeEtype}(b) at $F=0.6$ kV/cm. Such resonances are
realized for electron as well as hole tunneling. For this case the
molecular design results in significant oscillator strength for
both cascade transitions. In both cases we obtain an excellent fit
of the measured biexciton-exciton spectrum, as seen by the
comparison of Figs. \ref{fig:XXcascadeEtype}(a) and (b). Thus we
find a common qualitative understanding of both types of neutral
biexciton/exciton spectra in asymmetric QDMs.
\\
\indent We have presented the two-photon cascade spectra in QDMs
and shown that their energy levels are widely adjustable by
structural design and with electric field. Finally we note that
the biexciton and exciton in an uncharged QDM is only one example
of this class of transitions. We have also observed spectra for
singly (see dashed box Fig. \ref{fig:Overview}(a)) and doubly
charged QDMs, in which cases the transitions take place in the
presence of resident carriers. Extension of the current work to
these cases will provide the added opportunities of using
long-lifetime spin based quantum memories, and entanglement
between the photons and the resident carriers.
\begin{acknowledgments}
We acknowledge the financial support by NSA/ARO and ONR. E.A.S.,
I.V.P., and M.F.D. thank the NRC/NRL for financial support.
\end{acknowledgments}

\bibliography{LitXXpaper}

\begin{thebibliography}{15}
\bibitem[{PL-()}]{PL-triplets2}
\bibinfo{note}{At high excitation powers also triplets of PL-lines can be seen,
  corresponding to 0, 1, and 2 charges in the top QD.}
\bibitem[{Z-H()}]{Z-HamiltonX0-alt}
\bibinfo{note}{For hole or electron tunneling the molecular neutral exciton
  energies are found from the Hamiltonian with the basis spin states consisting
  of two bright, ($^{\uparrow0}_{\Downarrow0}$), ($^{\uparrow0}_{0\Downarrow}$)
  or ($^{0\uparrow}_{\Downarrow0}$), and two dark,
  ($^{\uparrow0}_{\Uparrow0}$), ($^{\uparrow0}_{0\Uparrow}$) or
  ($^{0\uparrow}_{\Uparrow0}$), excitons \begin{tiny}\begin{equation*}
  \label{eqn:X0Ham} \widehat{H}_X=\widehat{I}\varepsilon_X \\
  +\left(\begin{array}{cccc} J^{eh} & \frac{1}{2}\Delta^X & 0 & 0 \\
  \frac{1}{2}\Delta^X & E^{eh}+pF & 0 & 0 \\ 0 & 0 & -J^{eh} &
  \frac{1}{2}\Delta^X \\ 0 & 0 & \frac{1}{2}\Delta^X & E^{eh}+pF \\ \end{array}
  \right) \end{equation*}\end{tiny} The Hamiltonian for the biexciton singlets
  ($^{20}_{20}$), ($^{20}_{11}$)$_S$ or ($^{11}_{20}$)$_S$, ($^{20}_{02}$) or
  ($^{02}_{20}$), and one of the three degenerate and fully decoupled triplets
  ($^{20}_{11}$)$_T$ or ($^{11}_{20}$)$_T$ of the biexciton is
  \begin{tiny}\begin{equation*}
  \label{eqn:XXHam}\widehat{H}_{XX}=\widehat{I}\varepsilon_{XX}\\
  +\left(\begin{array}{cccc} 0 & \frac{\sqrt{2}}{2}\Delta^X & 0 & 0 \\
  \frac{\sqrt{2}}{2}\Delta^X & E^{h(e)}_{1}+pF & \frac{\sqrt{2}}{2}\Delta^X & 0
  \\ 0 & \frac{\sqrt{2}}{2}\Delta^X & E^{h(e)}_{2}+2pF & 0 \\ 0 & 0 & 0 &
  E^{h(e)}_{1}+pF \\ \end{array}\right) \end{equation*}\end{tiny} Here
  $\varepsilon_{X(XX)}$ is the confinement energy of an exciton(biexciton) in a
  single QD (here bottom QD), $\widehat{I}$ is an identity matrix and $E^{eh}$,
  $E_{1}^{h(e)}$ and $E_{2}^{h(e)}$ are the Coulomb energies that are required
  to move one or two holes(electrons) from the bottom to the top QD.}
\end{thebibliography}
\end{document}